\documentclass[pra, notitlepage, superscriptaddress,  reprint]{revtex4-1}
\usepackage{amssymb}
\usepackage{amsmath}
\usepackage{braket}
\usepackage{graphicx}
\usepackage[usenames,dvipsnames]{color}
\usepackage{tensor}
\usepackage{esint} 
\usepackage{empheq}
\usepackage{capt-of}
\usepackage[normalem]{ulem} 
\usepackage{soul} %

\definecolor{myorange}{RGB}{199.24, 87.48, 47.80}
\usepackage[colorlinks,bookmarks=false,citecolor=NavyBlue,linkcolor=OliveGreen,urlcolor=blue]{hyperref}

\newcommand{\nn}{\nonumber}

\newcommand{\be}{\begin{equation}}
\newcommand{\ee}{\end{equation}}
\newcommand{\ba}{\begin{aligned}}
\newcommand{\ea}{\end{aligned}}
\newcommand{\bw}{\begin{widetext}}
\newcommand{\ew}{\end{widetext}}

\newcommand{\pint}{{\rm P}\int}

\newcommand{\resatilde}{\tilde{a}_{\text{res}(h_i)}}
\newcommand{\phset}{\mathbf{p},\mathbf{h} }

\def\doi{http://dx.doi.org/}

\begin{document}
\title{Edge singularities and quasi-long-range order in non-equilibrium steady states}
\author{Jacopo De Nardis}
\affiliation{D\'epartement de Physique, \'Ecole Normale Sup\'erieure / PSL Research
University, CNRS, \\24 rue Lhomond, 75005 Paris, France}
\author{Mi\l osz Panfil}
\affiliation{Institute of Theoretical Physics, University of Warsaw,
 ul.~Pasteura 5, 02-093 Warsaw, Poland}
\begin{abstract}
The singularities of the dynamical response function are one of the most remarkable effects in many-body interacting systems. However in one dimension these divergences only exist strictly at zero temperature, making their observation very difficult in most cold atomic experimental settings. Moreover the presence of a finite temperature destroys another feature of one-dimensional quantum liquids: the real space quasi-long-range order in which the spatial correlation functions exhibit power-law decay. We consider a non-equilibrium protocol where two interacting Bose gases are prepared either at different temperatures or chemical potentials and then joined.  We show that the non-equilibrium steady state emerging at large times around the junction displays edge singularities in the response function and quasi-long-range order.
\end{abstract}
\maketitle

\textit{Introduction}. X-ray edge singularities are one of the most spectacular phenomena of strongly correlated fermionic systems. These are divergences (in general non-analycities) of the response functions in the vicinity of the threshold energies caused by the Fermi sea structure of the many-body ground state. The general theory of edge singularities was developed in late 60's~\cite{1967_Anderson_PRL_18,1967_Mahan_PR_163,1969_Nozieres_PR_178,1969_Schotte_PR_182,1969_Schotte_PR_185,1990_Ohtaka_RMP_62,Mahan_BOOK} and since then is one of the hallmarks of nonperturbative quantum many-body physics.

In metals, absorption of a high energy photon (X-ray) with momentum $k$ creates a core hole by exciting one of the electrons to the conductance band. At zero temperature the Fermi sea is completely filled and therefore there is a threshold energy $\omega_-(k)$ for such a process to occur. The response of the system is then controlled by two competing processes. The created core hole for the conductance electrons leads to the orthogonality catastrophe~\cite{1967_Anderson_PRL_18} which decreases the response. On the other hand, the attractive interaction between the electron and the core hole enhances the response~\cite{1967_Mahan_PR_163}. Both effects are nonperturbative and the result of their competition is encoded in the exponent $\mu(k)$ controlling the behavior of the dynamic structure factor (\textsc{dsf}) in the vicinity of $\omega_-(k)$
\begin{equation}
  S(k,\omega) \simeq  S_-(k) | {\omega - \omega_-(k)} |^{\mu(k)}.
\end{equation}
The threshold exponent $\mu(k)$ is proportional to the scattering phase of conduction electrons at the Fermi surface with the core hole. The scattering phase, that depends on the microscopic interactions, can be negative or positive resulting in either singularity or vanishing of the~\textsc{dsf}. Over the years the edge singularities were observed in many electronic systems. The most direct is the absorption of the X-rays in metals, e.g.~\cite{PhysRevB.13.2411,doi:10.1143/JPSJ.42.876,PhysRevB.18.6622}
. They appear also in other situations like, for example, a resonant tunneling experiments~\cite{1994_Geim_PRL_72} or a quantum dot coupled to a degenerate electron gas~\cite{2013_Haupt_PRB_88}. 

\begin{figure}[t]
  \center
 \includegraphics[width=0.45\textwidth]{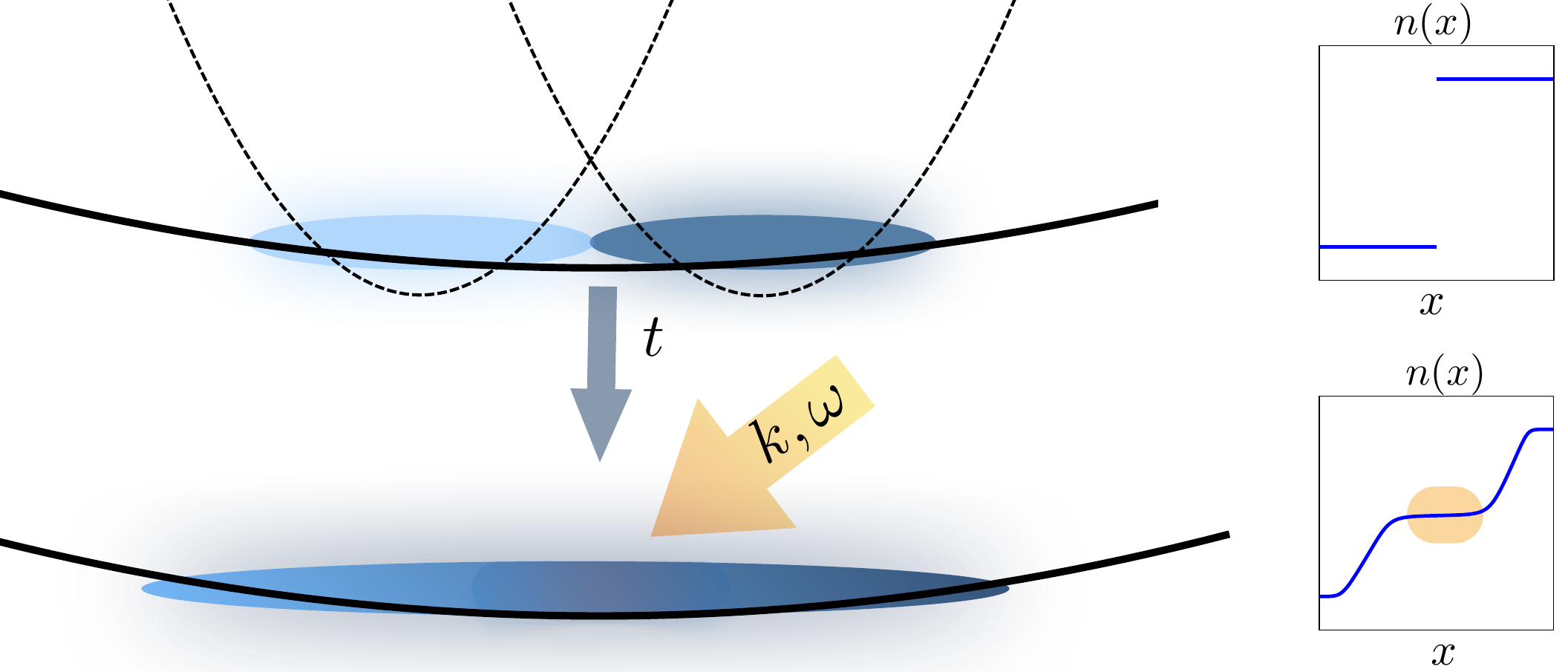}
  \caption{Non-equilibrium protocol to observe edge singularities in the \textsc{ness}. Top: two quantum gas at different density $n$ are prepared and then joined together, Bottom. After waiting some time that the system reaches its steady state close to the junction (see plot of the density $n(x)$ on the right) a Bragg pulse is shined around the center of the cloud to probe its dynamical correlations.}
  \label{fig:release}
\end{figure}

Phenomena of the same nature appear also in non-metallic one-dimensional (1d) systems. In 1d the presence of the interactions leads to a formation of the Fermi sea also for non-fermionic systems such as the Lieb-Liniger gas of bosons~\cite{1963_Lieb_PR_130_1}. The Fermi sea structure of the ground state is a universal feature of 1d quantum liquids as described by the Luttinger liquids (LL) theory~\cite{1981_Haldane_PRL_47,GiamarchiBOOK} which supersedes the Fermi liquid description valid in higher dimensions. The Luttinger liquid physics was very recently experimentally observed \cite{PhysRevLett.119.165701} and also in a number of other situations in the past years~\cite{Bockrath1999,physRevLett.102.107204,Auslaender825}. Like in metals, the presence of an effective Fermi sea and the interactions create suitable conditions for the appearance of the edge singularities. This intuition resulted in a full-fledged theory of non-linear Luttinger liquids~\cite{2012_Imambekov_RMP_84} (nLL), for which the Lieb-Liniger model served as the main playground~\cite{PhysRevLett.100.206805}. The theory of nLL predicts the edge singularities in the response functions to be a universal feature of the ground states of quantum liquids in 1d~\cite{RevModPhys.84.1253,PhysRevLett.100.027206,2009_Imambekov_SCIENCE_323,2008_Imambekov_PRL_100,
  2012_Imambekov_RMP_84,1367-2630-17-10-103003}.

However, the edge singularities in 1d so far escaped from experimental observation except for few qualitative results as in~\cite{Moreno2016}. The main reason being that the ground state physics of gapless systems is obscured by the usual presence of finite temperature fluctuations that hide the quantum correlations. Indeed, {while in 3d metals the Fermi temperature is of order of $10^3$ kelvins, for ultracold gases in typical experimental settings this is of order of nanokelvins.} On the other hand, the edge singularities physics is not limited to equilibrium states. Indeed the main ingredient necessary for their appearance is a discontinuity in the fermionic occupation number.
Non-equilibrium states of matter displaying edge singularities were theoretically proposed in past, such  
like a state with two or more Fermi seas with different chemical potentials~\cite{1996_Ng_PRB_54,2005_DAmbrumenil_PRB_2005,2005_Abanin_PRL_94}. Another example is a defected Fermi sea (Moses state) which was introduced both in 3d~\cite{2011_Bettelheim_JPA_44} and 1d systems~\cite{PhysRevA.89.033637,SciPostPhys.1.1.008}. The past years have witnesses huge developments in studies of low dimensional systems out of equilibrium \cite{RevModPhys.83.863,1742-5468-2016-6-064001,Eisert2015,DAlessio2016,1703.09740,LangenBOOK} and on the possibility of creating exotic states of matter via out-of-equilibrium protocols \cite{PhysRevLett.110.125302,PhysRevLett.115.175301,PhysRevB.93.115113,Diehl2011,Vestraetenature,PhysRevA.96.033828,Lange2017}. Among them interesting  from the edge singularities point of view are the so-called bi-partite quench protocols \cite{1742-5468-2008-11-P11003,1751-8121-45-36-362001,PhysRevB.89.214308,PhysRevA.91.021603,1751-8121-48-9-095002,1742-5468-2016-5-053108,SciPostPhys.3.3.020,PhysRevB.90.161101,SciPostPhys.1.2.014,PhysRevE.96.012138,SciPostPhys.2.1.002,PhysRevB.88.195129,PhysRevB.93.205121,PhysRevLett.119.020602,PhysRevB.95.045125,PhysRevB.90.155104,PhysRevB.95.045125,Ben,arXiv.1605.09790,PhysRevB.96.115124,1707.06218,1706.00020,PhysRevB.96.220302,PhysRevX.7.021012}. It consists of two extended and independent systems at thermal equilibrium albeit at different temperatures $T_L$ and $T_R$ and/or densities $n_L$ and $n_R$. At time $t=0$ the two systems are connected, see Fig. \ref{fig:release}, and at late times close to the junction a translationally invariant non-equilibrium steady state (\textsc{ness}) emerges. Given the (quasi-) particles of the model with momentum $k(\lambda)$ (with $\lambda \in \mathbb{R}$ the rapidity variable which parametrizes particle momenta
$k(\lambda)$) and their (dressed) dispersion relation $\varepsilon(\lambda)$ their momentum distribution $\vartheta(\lambda) \in [0,1]$ in the \textsc{ness} takes the following form \cite{Ben,arXiv.1605.09790} in terms of the Heaviside $\theta$ function
\begin{equation}\label{eq:NESSvartheta}
  \vartheta_{\text{\textsc{ness}}}(\lambda)= \theta( v_{\text{\textsc{ness}}}(\lambda) ) \vartheta_L (\lambda)+ \theta( -v_{\text{\textsc{ness}}}(\lambda) ) \vartheta_R(\lambda).
\end{equation}
Here the velocity $v_{\text{\textsc{ness}}}(\lambda)$ is the group velocity of the quasi particles in the \textsc{ness} state. For interacting models this is usually a functional of the distribution $\vartheta_{\text{\textsc{ness}}}(\lambda)$ and therefore equation \eqref{eq:NESSvartheta} should be solved recursively. While the \textsc{ness} state \eqref{eq:NESSvartheta} is only well defined for models with stable quasi-particle excitations (i.e. integrable models), recently it was shown that at low temperature the form \eqref{eq:NESSvartheta} is also valid for generic 1d interacting models \cite{1709.10096}.

The presence of the discontinuity in~\eqref{eq:NESSvartheta}, at $\lambda_0$ such that $v_{\rm \textsc{ness}}(\lambda_0)=0$, suggests, according to the general theory introduced above, that the \textsc{dsf} of the \textsc{ness} state might exhibit edge singularities. In this letter, we show that this is indeed the case and we characterize the threshold energies and threshold exponents. To achieve this we use a recently developed approach to the \textsc{dsf} for the Lieb-Liniger model introduced in~\cite{smooth_us,thermo_us}.

\textit{Dynamical correlations of the \textsc{ness}.} We focus here on the Lieb-Liniger model  \cite{1963_Lieb_PR_130_1} for interacting bosons with repulsive coupling $c>0$. The model is experimentally relevant for cold atomic physics  \cite{1998_Olshanii_PRL_81,2008_Amerongen_PRL_100,Bouchoule,PhysRevA.91.043617,PhysRevLett.115.085301,PhysRevA.96.033624,2017arXiv170800031P,1742-5468-2016-6-064009,Meinert945,PhysRevLett.113.035301} and  its non-equilibrium properties, especially after a quantum quench \cite{1712.04642,2012_Shashi_PRB_85,PhysRevA.89.013609,2014_DeNardis_PRA_89,PhysRevA.91.051602,1751-8121-48-43-43FT01,1367-2630-18-4-045010,
  1710.11615,1742-5468-2016-6-064009,Foini2017,1751-8121-50-50-505003,1712.05262,SciPostPhys.3.3.023} have attracted a large attention in the past years. Its Hamiltonian density is given by
\begin{equation}
  \mathcal{H}(x) = -\psi^{\dagger}(x)\partial_x^2\psi(x) + c \ \psi^{\dagger}(x)\psi^{\dagger}(x)\psi(x)\psi(x),
\end{equation}
with $\psi(x)$ the canonical Bose field. The group velocity of the quasi-particles is given in terms of $\vartheta$ by an integral equation \cite{SciPostPhys.2.2.014}
\begin{equation}
v(\lambda) = 2 \lambda  + \int_{-\infty}^{+\infty} \frac{d\mu}{2 \pi} k'(\mu)\theta(\lambda- \mu)  {  \vartheta(\mu)}{} (v(\mu)- v(\lambda)),
\end{equation}
with the scattering phase of the model $\theta(\lambda) = 2 \arctan(\lambda/c)$ and with the momentum given by $k(\lambda) = \lambda + \int d\mu \theta(\lambda-\mu) \vartheta(\mu)$.
We denote the height of the discontinuity in the particle distribution by $\Delta \vartheta = \vartheta_L(\lambda_0) - \vartheta_R(\lambda_0)$.
\begin{figure}[t]
  \center
 \includegraphics[width=0.47\textwidth]{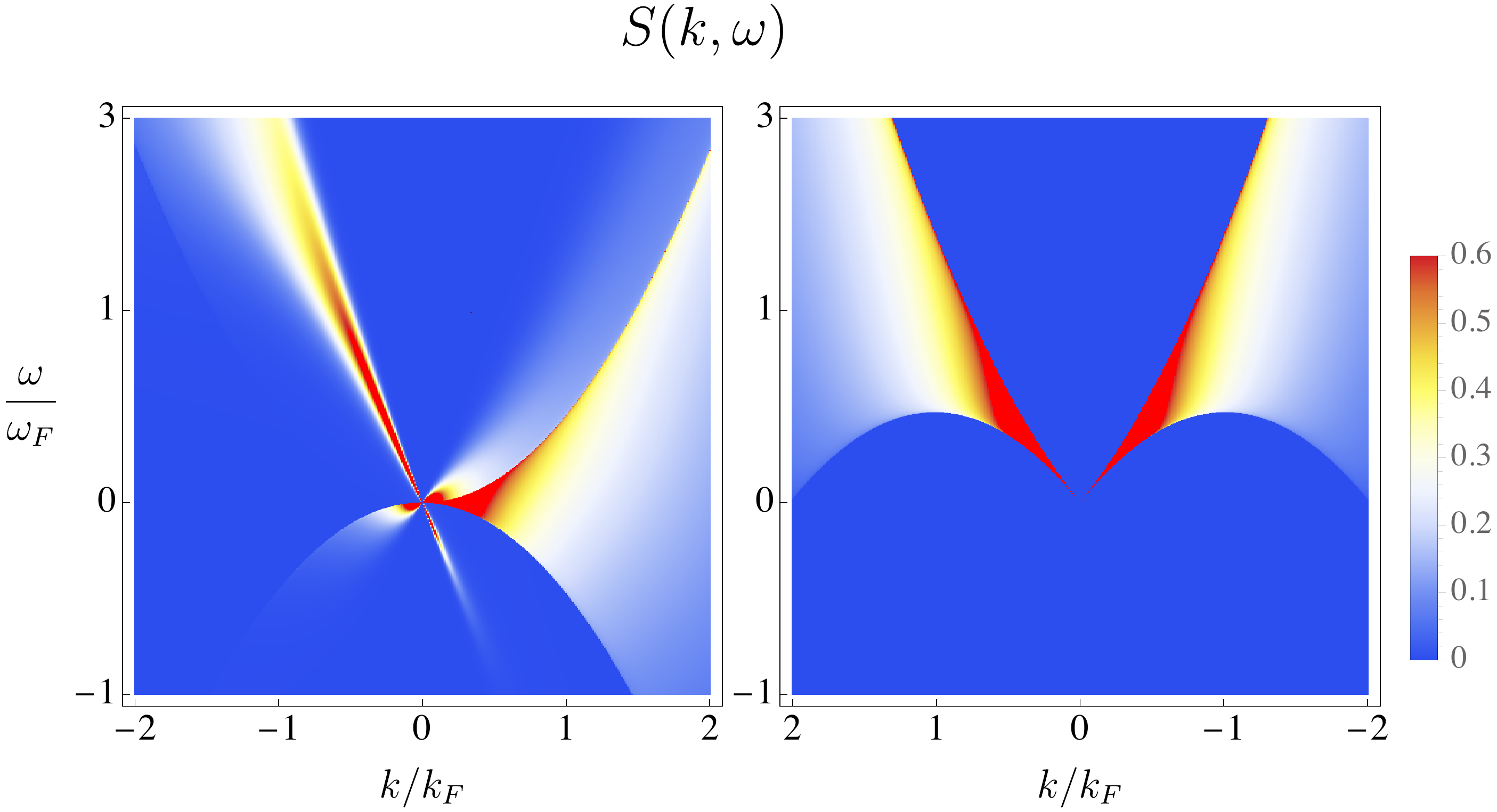}
 \caption{Density plot of the \textsc{dsf} (in unit of $n/\omega_F$) computed on the \textsc{ness} (left) and on the ground state (right) of a gas with density  $n=n_{\text{\textsc{ness}}}$, and coupling strength $c=10$ ($k_F = \pi n$, $\omega_F = k_F^2$). The \textsc{ness} is obtained by joining a left gas with $T_L=1,n_L=1$ and a right gas with $T_R=1,n_R= 0.1$. The coupling strength $c=10$ of both gases is the same  (density $n_{\text{\textsc{ness}}} = 0.54$).
   In the \textsc{ness}  and for $k>0$ the \textsc{dsf} is divergent close to the the particle excitations $\omega_{+}(k) \sim  {k^2}/{2 m_0}$ and it has a zero in proximity of the hole edge $\omega_{-}(k) \sim -  {k^2}/{2 m_0}$. For negative $k$ the situation is reversed. In the ground state instead the \textsc{dsf} has a pole (zero) around the particle (hole) edge $\omega_{\pm}(k) \sim v_s |k| \pm  {k^2}/{2m_F}  $, see for example \cite{2008_Imambekov_PRL_100,PhysRevA.74.031605}.} \label{fig:DSF_ALL}
\end{figure}

\begin{figure}[t]
 \centering
 \includegraphics[width=0.9\hsize]{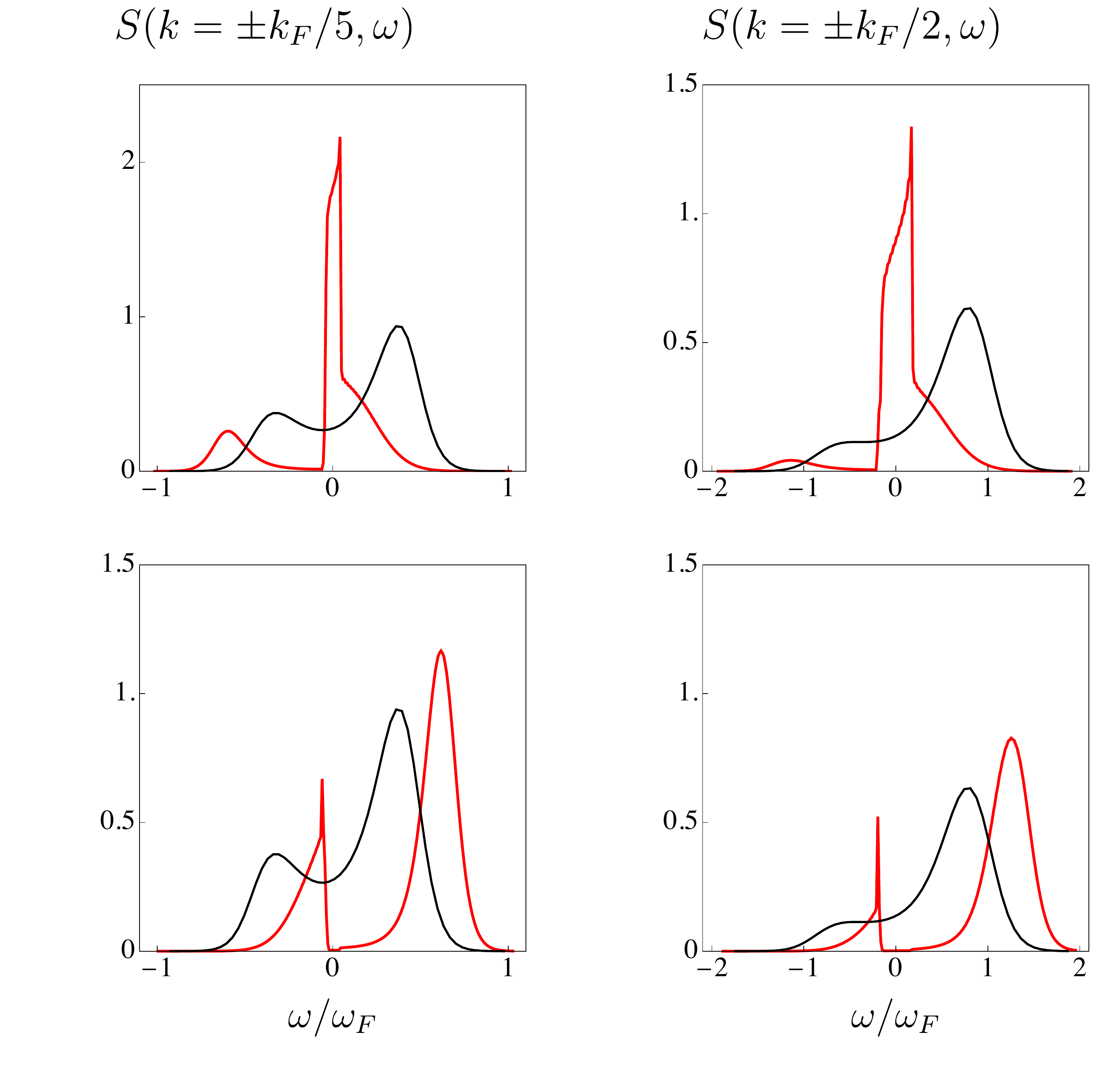} 
\caption{\textsc{dsf} \eqref{S_1ph_2} of the \textsc{ness} (red line) and on a thermal state with $T=1$, $n=n_{\text{\textsc{ness}}}$, and $c=4$. Plots on top display data with $k=k_F$ and the ones on the bottom with $k=-k_F$ ($k_F = \pi n $). The \textsc{ness} is the same as the one computed in Fig. \ref{fig:DSF_ALL}.  } \label{fig:DSF}
 \end{figure}
 
We consider the two-point correlation function $S(x,t) = \langle {\rm \textsc{ness}}|\hat{\rho}(x,t)\hat{\rho}(0,0)|{\rm \textsc{ness}}\rangle$ of the density operator $\hat{\rho}(x) = \psi^{\dagger}(x) \psi(x)$. Our main interest is its Fourier transform, the \textsc{dsf}
\begin{equation}\label{eq:DSF}
  S(k,\omega) = \int_{-\infty}^{\infty} {\rm d}t\int_{-\infty}^\infty {\rm d}x e^{i(kx-\omega t)}S(x,t).
\end{equation}
In the Lieb-Liniger gas, being an interacting model, the density operator can create any number of pairs of particle-hole excitations on the reference state. Each particle-hole pair corresponds to a local (small) modification of the filling function $\vartheta(\lambda)$ at positions of the particle $p$ and hole $h$. {We denote a state with a single particle-hole pair as} $|{\rm \textsc{ness}} , h \to p \rangle$. The momentum $k$ and the energy $\omega$ of this excited state with respect to the $|{\rm \textsc{ness}}\rangle$ is \cite{KorepinBOOK}
\begin{equation}
  k = k(p) - k(h),\qquad
  \omega = \varepsilon(p) - \varepsilon(h), \label{ph_kw}
\end{equation}
where $k(\lambda)$ was defined above and $\varepsilon(\lambda)$ is the dressed energy
$\varepsilon(\lambda) = \lambda^2 + 2\int_{-\infty}^{\infty}{\rm d}\alpha\ \alpha F(\alpha|\lambda)\vartheta(\alpha)$
such that the group velocity is given by $v(\lambda)= \partial \varepsilon(\lambda) / \partial k (\lambda)$. 
The back-flow function (the dressed scattering phase) obeys an integral equation
\begin{equation}
  2\pi F(\lambda|\alpha) \!=\! \theta(  \lambda - \alpha) +\! \int_{-\infty}^{\infty}\!\!{\rm d}\lambda' \vartheta(\lambda')F(\lambda'|\alpha)K(\lambda - \lambda'),
\end{equation}
with the scattering kernel given by $K(\lambda) = d \theta(\lambda) /d\lambda$. 

The spectral representation of $S(k,\omega)$ can be organized in the sum over number of created particle-hole excitations and
expressed through the form factor of the density operator 
$\langle {\rm \textsc{ness}}|\hat{\rho}(0)|{\rm \textsc{ness}} , h \to p  \rangle$. These form factors for the Lieb-Liniger model on a finite line of length $L$ were derived in~\cite{1990_Slavnov_TMP_82} and since then studied and used in the computation of the correlation functions \cite{1742-5468-2011-12-P12010,1742-5468-2012-09-P09001,2011_Kozlowski_JSTAT_P03019,2012_Shashi_PRB_85_1}. In~\cite{smooth_us,SciPostPhys.1.2.015,thermo_us} we have studied a general expression for the thermodynamic limit of such form factors and shown that their form depends strongly on the analyticity of the distribution function $\vartheta(\lambda)$, see~\footnote{See Supplemental Material}. One way the singular behavior presents itself is in the fractional dependence of the form factor with the system size $L$. This signals a need for a resummation (dressing) of the form factor as was for example demonstrated in~\cite{1742-5468-2011-12-P12010,1742-5468-2012-09-P09001}. However the dressing is only important at higher momenta and as shown in~\cite{thermo_us} there is a discontinuity dependent cutoff momentum $k^*= \left|\frac{k'(\lambda_0)}{\partial_{\lambda}F(\lambda_0|\lambda)|_{\lambda=\lambda_0}}\right|$ below which the dressing is not important. Staying below this cutoff the \textsc{dsf} has a standard expansion in momentum: the $m$ particle-hole pairs contribute at order $k^{m-1}$ to the \textsc{dsf}. Therefore the leading order at small momentum~\footnote{The value of $k$ should be compared with other scale of dimension inverse length which is $c$. Therefore the small momentum limit means $k/c$ is small or introducing a dimensionless coupling constant $\gamma = c/n$ the small momentum limit means $(k/k_F) (\pi/\gamma)$ is small. The Fermi momentum is $k_F = \pi n$. } is given by the form factors of a single particle-hole pair~
~\cite{thermo_us}
\begin{equation}
  |\langle {\rm \textsc{ness}}|\hat{\rho}(0)|{\rm \textsc{ness}} , h \to p \rangle| = k'(p) \left|\frac{\lambda_0 - h}{\lambda_0 - p}\right|^{\Delta\vartheta F(\lambda_0)} .
\end{equation}
Here $F(\lambda) = F(\lambda|p) - F(\lambda|h)$ is the back-flow of the particle-hole excitation and its sign is opposite to  that of the momentum $k$. 

For each choice of $(k,\omega)$ there is the corresponding excitation $(p,h)$. The correlation function in~\eqref{eq:DSF} is then equal to a single form factor contribution multiplied by the Jacobian of the change of variables from $(p,h)$ to $(k,\omega)$ and up to corrections of order $k/c$ is~\cite{Note1}
\begin{equation}
  S(k,\omega) = \mathcal{D}(k,\omega) |\langle {\rm \textsc{ness}}|\hat{\rho}(0)|{\rm \textsc{ness}} , h \to p \rangle|^2. \label{S_1ph_2}
\end{equation}
The density of states $\mathcal{D}(k,\omega)  = \frac{   \vartheta(h)  ( 1- \vartheta(p) ) }{| v(p) - v(h)|}$ and the position of particle and hole $p,h$ are given by the energy and momentum conservation. 
The same formula, with $|{\rm \textsc{ness}}\rangle$ replaced by a thermal state~\cite{KorepinBOOK,PhysRevA.89.033605}, holds at thermal equilibrium. Figures~\ref{fig:DSF_ALL} and \ref{fig:DSF} show the dynamic structure factor computed with this formula for the \textsc{ness} state and a thermal state.

The singular behavior of the \textsc{dsf} on the \textsc{ness} is similar to the one encountered for the ground state of the Lieb-Liniger model. There, the Fermi sea structure $\vartheta_{\text{GS}}(\lambda) = \theta(\lambda + \lambda_F) \theta(\lambda_F-\lambda)$ (with $\lambda_F$ the Fermi rapidity such that $k(\pm \lambda_F) = \pm k_F$) leads to two fundamental types of the particle-hole excitations~\cite{1963_Lieb_PR_130_2}. The particle type describes excitation in which a hole is created at the edge of the Fermi sea and the particle is free to move. The hole type corresponds to a reversed situation when the particle position is fixed to the edge of the Fermi sea and the hole instead is free to move. The form factors of the density operator are singular for both types of excitations which leads to well-known singularities of the correlation functions as universally described by the non-linear Luttinger liquid~\cite{2008_Imambekov_PRL_100,2009_Imambekov_SCIENCE_323,2012_Imambekov_RMP_84}.

\begin{figure}[t]
  \center
 \includegraphics[width=0.9\hsize]{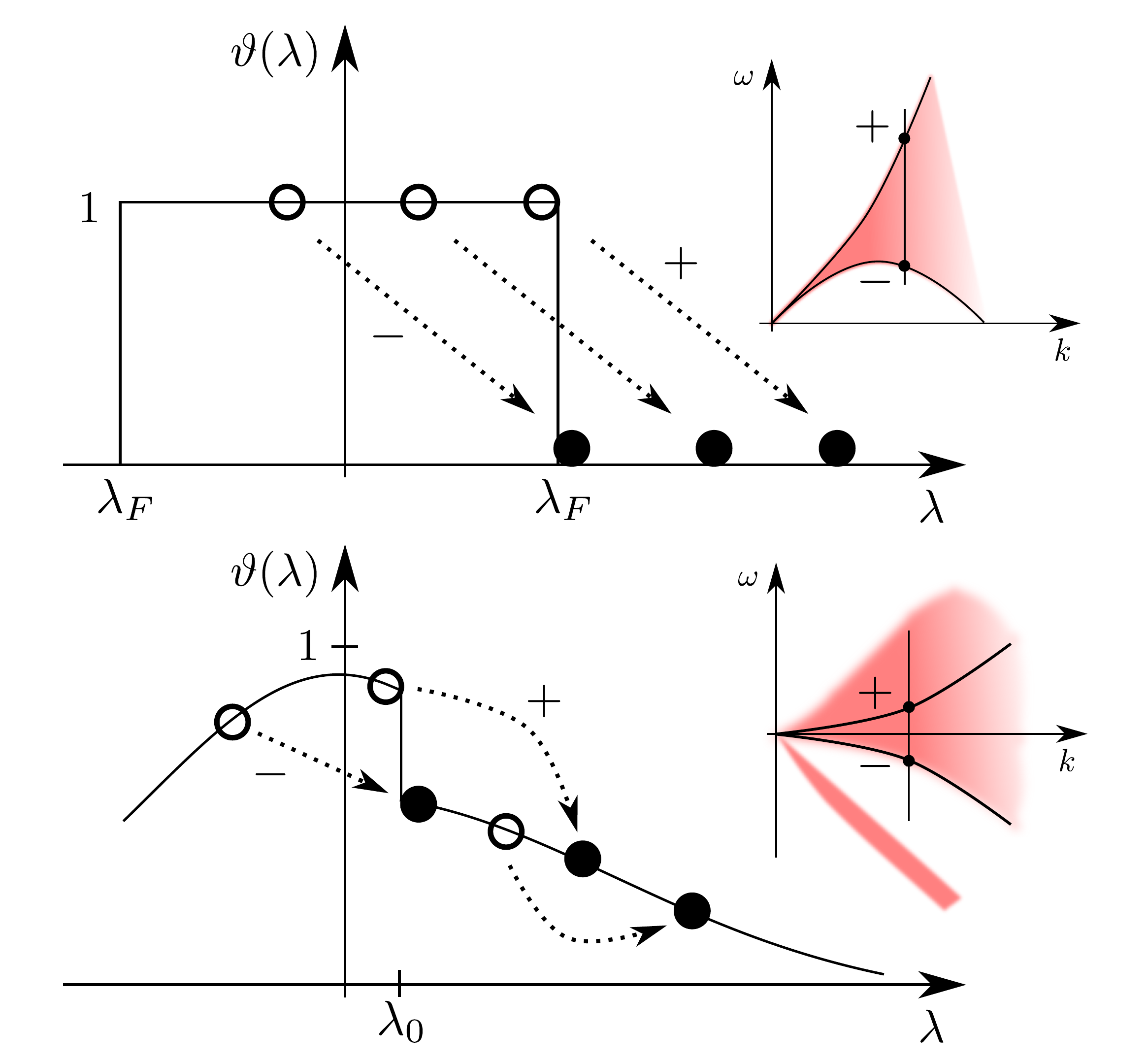}
  \caption{The distribution function $\vartheta(\lambda)$ for the ground state (top) and the \textsc{ness} state (bottom). For a fixed momentum $k$ a particle-hole excitation of the ground state can be created only for a finite interval of energies $\omega_+(k) - \omega_-(k)$. The edge excitations correspond to a configuration where either particle or hole is right at the edge of the Fermi sea $\lambda_F$. On the contrary, over the \textsc{ness} state, particle-hole pairs of arbitrary energy can be created however there are again two edge modes where either particle or hole is placed at the discontinuity $\lambda_0$. }
  \label{fig:edges}
\end{figure}

Here we face a similar situation, see Fig. \ref{fig:DSF_ALL}, \ref{fig:edges}. However there are  differences. First, there is only one discontinuity, at $\lambda_0$. Second the \textsc{ness} distribution is asymmetric which implies that $S(k,\omega)$ is not a symmetric function of $k$, see Fig.~\ref{fig:DSF_ALL}.
The third difference is related to the nature of the excitations. The ground state Fermi sea is completely filled up to $\lambda_F$ and then empty. This means that the hole, which for the particle excitations, should be placed at the edge can be only placed just before the edge. For the \textsc{ness} the situation is different, the distribution function is neither $1$ or $0$ in the vicinity of $\lambda_0$. Therefore the hole can be created on both sides of the discontinuity. The situation is analogous to the type II excitation where also particle can be placed on both sides of the discontinuity. Therefore the particle and hole excitations are themselves formed by two different microscopic configurations with the same dispersion relations.

The dispersion relations for the particle and hole excitations can be derived with a standard thermodynamic Bethe ansatz techniques~\cite{KorepinBOOK}. At small $k$ they read 
\begin{equation}
\omega_{\pm}(k) =\pm \frac{k^2}{2 m_0}   + O(k^3), \quad m_0 = \frac{k'(\lambda_0)}{v'(\lambda_0) }.
\end{equation}
While the ground state of quantum 1d liquids supports sound waves excitations, excitations in the \textsc{ness} around the edge are massive since the velocity $v(\lambda)$ vanishes at~$\lambda_0$.

The singularities in the \textsc{DSF} appear when the energy $\omega$ is close to $\omega_{\pm}(k)$. Explicitly, formula~\eqref{S_1ph_2}, in the vicinity of either singularity, is
\begin{equation}
  S(k,\omega) \simeq  \  S_{\pm}(k) \left|\frac{\omega - \omega_-(k)}{\omega - \omega_+(k)}\right|^{\mu(k)},
\end{equation}
with the exponent $\mu(k)$ and the momentum dependent prefactors $S_{\pm}(k)$ given in their leading order in $k$ by
\begin{align}
  \mu(k) &= -2 \Delta \vartheta F(\lambda_0) \label{threshold_exp_small_k},\\
  S_{+}(k) &= \frac{m_0 k'(\lambda_0)^2}{|k|} (\vartheta_L(\lambda_0) + \vartheta_R(\lambda_0))(1 - \vartheta_R(\lambda_0)), \nonumber \\
  S_{-}(k) &= \frac{m_0k'(\lambda_0)^2}{|k|} \vartheta_L(\lambda_0)(2 - (\vartheta_L(\lambda_0) + \vartheta_R(\lambda_0))).
\end{align}
The sum of the distribution functions in $S_{\pm}(k)$ reflects the aforementioned fact that each mode is made of two microscopic types of configurations. The singularity itself is controlled by the exponent $\mu(k)$. For the \textsc{ness} state of Fig.~\ref{fig:release} it results in a divergence along the particle excitations at positive $k$ and along hole excitations for negative $k$. 

\textit{Spatial correlations.} The presence of a discontinuity in the occupation number $\vartheta(\lambda)$ has also important consequences on the structure of the spatial density-density correlations. Static correlations in real space $S(x,0)$ can be expanded as a sum over particle-hole form factors, weighted by the momentum phase $e^{i (k(p)- k(h))x}$. For a \textsc{ness} state at large $x$ the sum over particle-hole position accumulates around the discontinuity and leads to a {power law decay of the correlations as
\begin{equation}
  \frac{S(x,0)}{n_{\text{\textsc{ness}}}^2}  \simeq 1 -  \frac{A_0}{2(\pi x n_{\text{\textsc{ness}}} )^2} + O(e^{-\beta_{L,R} |x| }), \quad x \gg n_{\text{\textsc{ness}}}^{-1} .
\end{equation}
 The amplitude $A_0$ is given by the matrix element of a single particle-hole excitation close to the discontinuity $\lambda_0$, namely $\lim_{p,h \to \lambda_0}| \langle \text{\textsc{ness}} | \hat{\rho} | \text{\textsc{ness}}, h \to p \rangle |^2 =  k'(\lambda_0)^2$, see \cite{SciPostPhys.1.2.015}. We obtain $A_0 = (k'(\lambda_0) \Delta \vartheta)^2/2 $.} Notice that this expression gives back the Luttinger liquid parameter $K = (k'(\pm\lambda_F))^2$ when the state is the ground state, see \cite{KorepinBOOK,2012_Shashi_PRB_85_1,1742-5468-2012-09-P09001}. 
 This shows the \textsc{ness} has much longer range density-density correlations compared to the left and right state, where the decay is instead exponential at large distances. {This effect is similar to the the dynamical quasicondensation of hard-core bosons observed in trap-release experiments \cite{PhysRevLett.93.230404,PhysRevLett.115.175301}.}

 \textit{Beyond small $k$.}
{Until now we have been considering the structure of singularities at small momenta. Comparing the obtained results with the nLL theory we can conjecture a formula for the edge exponents at arbitrary $k$.} 
The nLL theory predicts the threshold exponents of the ground state of the Lieb-Liniger model to be $1 - \mu_L(k) - \mu_R(k)$ with $\mu_{L(R)}(k) = (1 + F(\pm q))^2$
where $L(R)$ are contributions coming from the left and the right Fermi edges, both of height $1$. The presence of a non-trivial $\vartheta$ affects the scattering phase as $F(\lambda) \to \vartheta(\lambda)F(\lambda)$,  we conjecture the threshold exponents for the \textsc{ness} to be
\begin{equation}
  \mu(k) = 1 - (1 + \Delta\vartheta F(\lambda_0))^2.
\end{equation}
at any $k$.
In the small momentum limit the back-flow is small and we recover the threshold exponent~\eqref{threshold_exp_small_k}. In order to prove such a statement, one would need to formulate a nLL field theory for the excitations around the \textsc{ness}, which is currently not known. Certain progresses in this direction were recently reported on inhomogenous Luttinger liquids \cite{SciPostPhys.3.3.019,1712.05262}.


\textit{Conclusions:}
We have shown that the bi-partite non-equilibrium protocol leads to excited states with unusual properties. They have finite energy density and entropy like thermal states but despite that, they display correlations which are typical of the ground state, i.e. they exhibit edge singularities and quasi-long-range order. We considered here  an integrable model, but a \textsc{ness} has been shown to exist for any model described by a Conformal Field Theory \cite{1742-5468-2016-6-064005,1751-8121-45-36-362001} or strongly interacting theories in higher dimensions~\cite{Bhaseen2015}. Moreover, a \textsc{ness} should appear at intermediate time scales for any interacting theory sufficiently close to an critical \cite{PhysRevLett.119.110201,1709.10096} or integrable point~\cite{1742-5468-2016-6-064009}.
We believe that our results pave the way towards a field theoretical, universal, description of the \textsc{ness}, similarly to the non-linear Luttinger Liquid theory for the ground state~\cite{2009_Imambekov_SCIENCE_323} and general zero-entropy states \cite{1751-8121-49-49-495203}.

\textit{Acknowledgments:}
The authors would like to thank Sebas Eli{\"e}ns for important comments on the early version of the manuscript.
JDN acknowledges Maurizio Fagotti for numerous discussions. The authors acknowledge support from LabEx ENS-ICFP:ANR-10-LABX-0010/ANR-10-IDEX-0001-02 PSL* (JDN) and from the NCN under FUGA grant \mbox{2015/16/S/ST2/00448} (MP). 

\bibliography{Edges_NESS}

\onecolumngrid

\begin{center}
{\large{\bf Supplemental Material for\\ ``Edge singularities and quasi-long-range order in non-equilibrium steady states''}}
\end{center}

\section{Lieb-Liniger Bose gas and its dynamical structure factor in the thermodynamic limit}
We consider the Lieb-Liniger model on a finite but very long system of length $L$ with periodic boundary conditions. Eigenstates of the model are parametrized by a set of $N$ quasi-momenta or rapidities $\lambda_j$ which solve the Bethe ansatz equations~\eqref{bethe} and which parametrize the momentum $k(\lambda)$. In the thermodynamic limit $N,L \to \infty$ we can introduce the filling function $\vartheta(\lambda)$ taking values in $[0,1]$ and charachterizing the quasi-momentum distribution of the state.   We give few examples of the filling function for physically interesting states. The filling function of the ground state corresponds to a Fermi~sea
\begin{equation}
  \vartheta(\lambda) = \begin{cases}
    1, \quad -q \geq \lambda \geq q,\\
    0, \quad \textrm{otherwise}.
  \end{cases}
\end{equation}
The filling function for the finite temperature state \cite{1969_Yang_JMP_10} is
\begin{equation}
 \vartheta(\lambda) = \frac{1}{1 + \exp(\beta (\varepsilon(\lambda) - \mu))},
\end{equation}
where $\varepsilon(\lambda)$ is the dressed energy \eqref{drenergy} and $\mu$ the chemical potential. Other distributions of interest are the generalized Gibbs ensemble (GGE) states, with $\beta(\lambda)$ a positive function~\cite{PhysRevB.84.212404,SciPostPhys.3.3.023}
\begin{equation}
  \vartheta(\lambda) = \frac{1}{1 + \exp(\beta(\lambda) (\varepsilon(\lambda) - \mu))},
\end{equation}
Contrary to the ground state, the filling function for finite temperature and GGE states is a smooth function of $\lambda$ for any $\beta(\lambda)<\infty$. 

The excited states around the thermodynamic state $|\vartheta\rangle$ are created by making a number of particle-hole pairs in the filling function. An $m$ particle-hole excited state we denote $|\vartheta; \mathbf{p},\mathbf{h}\rangle$, where $\mathbf{p} = \{ p_j\}_{j=1}^m$ and $\mathbf{h} = \{ h_j \}_{j=1}^m$.  Sets $\mathbf{p}$ and $\mathbf{h}$  specify the particle-hole content of the excited state. There exist also more general excited states with different number of particles and holes but the form factors of the density operators vanishes for such states. An excited state has relative ({with respect to $|\vartheta\rangle$}) energy and momentum given by
\begin{align}
 k(\vartheta; \mathbf{p}, \mathbf{h}) &= \sum_{j=1}^m k(p_j) -  k(h_j),\label{k}\\
  \varepsilon(\vartheta; \mathbf{p}, \mathbf{h}) &= \sum_{j=1}^m \varepsilon(p_j) - \varepsilon(h_j).\label{omega}
\end{align}
 {The functions $k(\lambda)$ and $\varepsilon(\lambda)$ are the dressed momentum and energy and are given by} 
\begin{align}
k(\lambda) &= \lambda + \int_{-\infty}^{\infty} {\rm d}\alpha F(\alpha|\lambda) \vartheta(\alpha), \\
\varepsilon(\lambda) &= \lambda^2 + \int_{-\infty}^{\infty} {\rm d}\alpha F(\alpha|\lambda) \vartheta(\alpha)  (2 \alpha).\label{drenergy}
\end{align}
where $F(\alpha| \lambda )$ is the backflow or phase shift. 
In this work we are concerned with the density-density correlation functions, also known as a dynamic structure factor, DSF, in the thermodynamic limit at fixed total density $n$. The density operator is $ \hat{\rho}(x) = \psi^{\dagger}(x)\psi(x)$ and its time evolution in the Heisenberg picture is given by the Lieb-Liniger Hamiltonian
$
  \hat{\rho}(x,t) = e^{iHt}\hat{\rho}(x) e^{-iHt}.
$
The dynamic structure factor, is given by
\begin{equation}
  S(k,\omega) =  \int_{-\infty}^{\infty} {\rm d}x \int_{-\infty}^{\infty} {\rm d}t\, e^{i(kx - \omega t)} S(x,t).
\end{equation}
and it
can be written in the thermodynamic limit as a sum over a generic number of pairs of particle-hole excitations on the reference state $|\vartheta\rangle$
\begin{equation}\label{S_expansion}
  S(k,\omega) = \sum_{m \geq 1 } S^{m\text{ph}}(k,\omega)   ,
\end{equation}
where the contribution from $m$ particle-hole pairs is given by
\begin{equation} \label{mph_contribution}
 S^{\text{mph}}(k,\omega) =     \frac{(2\pi)^2}{(m!)^2}  \fint_{-\infty}^{\infty} {\rm d}{\mathbf p}_{m} {\rm d}{\mathbf h}_{m} |\langle \vartheta| \hat{\rho}(0)|\vartheta, \mathbf{h} \to  \mathbf{p}  \rangle|^2 \delta(k - k(\mathbf{p}, \mathbf{h}))\delta(\omega - \omega(\mathbf{p}, \mathbf{h})).
\end{equation}
Here the integration measure is defined as
\begin{equation} 
  {\rm d}\mathbf{p}_m{\rm d}\mathbf{h}_m = \prod_{j=1}^m {\rm d}p_j {\rm d}h_j\,  k'(p_j) k'(h_j) \vartheta(h_j) ( 1-  \vartheta(p_j)),
\end{equation}
and the finite part integral is defined as
\begin{equation} \label{finite_part_integral}
  \fint_{-\infty}^{\infty} {\rm d}h f(h) = \lim_{\epsilon\rightarrow 0^+}\int_{-\infty}^{\infty} {\rm d}h\,f(h+i\epsilon) + \pi i \underset{h=p}{\rm res} f(h).
\end{equation}
The finite part integral appears because the thermodynamic form factors $ |\langle \vartheta| \hat{\rho}(0)|\vartheta, \mathbf{h} \to  \mathbf{p}  \rangle|$ display {kinematic poles} (single poles) when $p_j$ coincides with $h_k$. The contribution from single particle-hole pair is
\begin{equation}
  S^{1\rm{ph}}(k,\omega) = (2\pi)^2 \int_{-\infty}^{\infty} {\rm d}p\,{\rm d}h |\langle \vartheta| \hat{\rho}(0)|\vartheta, h \to  p  \rangle|^2 \delta(k - (k(p) - k(h)))\delta(\omega - (\epsilon(p) - \epsilon(h))).
\end{equation}
Evaluating the integrals leads to a simple formula
\begin{equation}
  S^{1\rm{ph}}(k,\omega) = \mathcal{D}(k,\omega) |\langle \vartheta| \hat{\rho}(0)|\vartheta, h \to  p  \rangle|^2,
\end{equation}
where
\begin{equation}
  \mathcal{D}(k,\omega) = k'(p) k'(h) \vartheta(h) ( 1-  \vartheta(p)) \left|\det \frac{\partial(p,h)}{\partial(k,\omega)}\right|.
\end{equation}
is the density of states. The last part is the Jacobian of the change of variables from positions of the particle and hole to the corresponding  momentum and energy.  We can rewrite this in the following simple form in terms of the quasi-particle velocity $v(\lambda)$ 
\begin{equation}
  \mathcal{D}(k,\omega) = \frac{\vartheta(h)(1-\vartheta(p))}{|v(p) - v(h)|}.
\end{equation}

\section{Dressing equations } \label{sec:TL}
Given the scattering kernel dressed by the distribution $\vartheta(\lambda)$
\begin{equation}
K_\vartheta(\lambda,\mu) = K (\lambda- \mu) \frac{\vartheta(\mu)}{2\pi}
\end{equation}
we define its resolvent as its inverse kernel (where multiplications should be seen as matrix products on the continuum)
\begin{equation}
(\boldsymbol{1} + \hat{L}_{\vartheta})\left(\boldsymbol{1} - \hat{K}_{\vartheta} \right) = \boldsymbol{1}  = \left(\boldsymbol{1} - \hat{K}_{\vartheta} \right)(\boldsymbol{1} + \hat{L}_{\vartheta}),
\end{equation}
(the  {}{operator} $\boldsymbol{1}$ represents the identity $\boldsymbol{1}(\lambda,\lambda') = \delta(\lambda-\lambda')$)
which can be also expressed as the solution of the following integral equation
\begin{equation}  \label{L_int_eq}
  \hat{L}_{\vartheta}(\lambda, \lambda') =  \frac{\vartheta(\lambda')}{2\pi} \left(  K (\lambda - \lambda')  + \int_{-\infty}^{\infty} {\rm d}\alpha \hat{L}_{\vartheta}(\lambda, \alpha)  K(\alpha- \lambda')  \right).
\end{equation}
Moreover the resolvent is also proportional to the derivative of the shift function, namely
\begin{equation}\label{L-F}
\hat{L}_{\vartheta}(\mu,\lambda) = -\vartheta(\lambda) \partial_\mu F(\lambda|\mu).
\end{equation}
Then the derivative of the dressed energy and momentum can be expressed as application of the matrix $ (\boldsymbol{1} + \hat{L}_{\vartheta})$ to a vector $\vec{w}\equiv w(\lambda)$ on the continuum as 
\begin{align}
\vec{k'}  &=  (\boldsymbol{1} + \hat{L}_{\vartheta})  \vec{1}  , \\
\vec{\epsilon' } &=  (\boldsymbol{1} + \hat{L}_{\vartheta})  \vec{(2 \lambda)}.\label{drenergy}
\end{align}
Notice that $\vec{k'}$ is the dressing of the unity vector $\vec{1}$ and that is also why it can be denoted as dressed density $\vec{k}'\equiv \vec{1}^{\text{dr}}$. In general given a generic conserved operator $Q$ with single particle eigenvalue $q(\lambda)\equiv \vec{q}$, we can define the dressed eigenvalue as 
\begin{equation}\label{eqdressing}
\vec{q^{\text{dr}}} =  (\boldsymbol{1} + \hat{L}_{\vartheta})  \vec{q}.
\end{equation}

\section{Entropy of states and definition of thermodynamic form factors}\label{sec:entropyFF}
In order to define the thermodynamic form factors $\langle \vartheta|\hat{\rho}(0)|\vartheta; \mathbf{h}\rightarrow \mathbf{p}\rangle$ we consider a finite system with periodic boundary conditions. The eigenstates of the Hamiltonian are parametrized by a set of quantum numbers $\{I_j\}_{j=1}^N$, where $N$ is the number of particles, which map to the rapidities via the Bethe equations
\begin{equation} \label{bethe}
  \lambda_j = \frac{2\pi}{L} I_j + \sum_{k \neq j = 1}^N \theta(\lambda_j - \lambda_k), \quad \quad j=1,\ldots, N.
\end{equation}  
In the thermodynamic limit $N,L\rightarrow \infty$ with $N/L=n$ fixed, the rapidities get dense on the real line and therefore their position can be parametrized in terms of the filling $\vartheta(\lambda)$ which gives the ratio between the number of occupied rapidites $\lambda$ and the maximal allowed number, in the interval $[\lambda, \lambda+d\lambda)$. This specifies a thermodynamic state $|\vartheta\rangle$.  Let $\{I_j\}_{j=1}^N$ be a set of quantum numbers specifying a Bethe state such that in the thermodynamic limit its filling function is given by $\vartheta(\lambda)$. There are many choices of quantum numbers leading to the same filling function, thus to the same thermodynamic state $|\vartheta\rangle$. Their number is $\exp S[\vartheta]$ where $S[\vartheta]$ is the extensive Yang-Yang entropy~\cite{1969_Yang_JMP_10}
\begin{equation}\label{yangyang}
S[\vartheta] = L \int_{-\infty}^{\infty} {\rm d}\lambda k'(\lambda) \Big[  (1-\vartheta(\lambda))|\log(1-\vartheta(\lambda))| + \vartheta(\lambda) |\log \vartheta(\lambda) |\Big].
\end{equation}
We define a normalized thermodynamic state as
\begin{equation}
 |\vartheta\rangle = \lim_{N,L \to \infty} \exp\left(- \frac{1}{2} S[\vartheta] \right)\sum_{\{I_j\}} |\{I_j\}\rangle,
\end{equation} 
where the summation is over all the $e^{S[\vartheta]}$ microscopic states with the same $\vartheta(\lambda)$ in the thermodynamic limit.
We then define its thermodynamic form factors as
\begin{equation}
  \langle \vartheta|\hat{\rho}(0)|\vartheta,  \mathbf{h} \to \mathbf{p} \rangle = \exp\left(\frac{1}{2}\delta S[\vartheta, \phset]\right) \lim_{L,N\to\infty} L^{m}\langle\{ I_j^0\}|\hat{\rho}(0)|\{ I_j^0+\text{ph}\}\rangle.
\end{equation}
with the differential entropy is defined as the entropy of the excited state minus the one of the reference state
$
  \delta S[\vartheta, \phset] = S[\vartheta,\phset] - S[\vartheta].
$
The state $|\{I_j^0\}\rangle$ is called the averaging state and can be any state described by the filling function $\vartheta(\lambda)$ in the thermodynamic limit.  

\subsection{Thermodynamic form factors}
We now introduce a set of particles and relative holes $\phset = \{ p_i\}_{i=1}^m, \{h_i\}_{i=1}^m$. We define the following dressing function (with $\pint$ denoting the principal value integration)
\begin{equation}
  \tilde{a}^{[\mathbf{p}, \mathbf{h}]}(\lambda)  =  \frac{\sin[ \pi \vartheta(\lambda) F(\lambda)]}{2\pi \sin [ \pi F(\lambda)]} \prod_{k=1}^m   \left(  \frac{  p_k - \lambda}{ h_k - \lambda}
 \sqrt{\frac{K(p_k -\lambda)}{K(h_k - \lambda)} }\right)
e^{ -\frac{c}{2} \pint_{-\infty}^{\infty} {\rm d}
    \lambda'  \frac{\vartheta(\lambda')F(\lambda') K(\lambda' - \lambda)}{\lambda' - \lambda} },\label{atilde}
\end{equation}
which defines a new scattering kernel
\begin{equation}
 K^{[\mathbf{p}, \mathbf{h}]}(\lambda,\lambda') = K(\lambda,\lambda') \tilde{a}^{[\mathbf{p}, \mathbf{h}]}(\lambda').
\end{equation}
The resolvent of this new kernel is defined through
\begin{equation} \label{L_ph_def}
  \left(\boldsymbol{1}+L^{[\mathbf{p}, \mathbf{h}]}\right)\left(\boldsymbol{1} -K^{[\mathbf{p}, \mathbf{h}]}\right) = \boldsymbol{1} = \left(\boldsymbol{1} -K^{[\mathbf{p}, \mathbf{h}]}\right)\left(\boldsymbol{1} + L^{[\mathbf{p}, \mathbf{h}]}\right).
\end{equation}
which can also be expressed via an integral equation 
\begin{equation}
  L^{[\mathbf{p}, \mathbf{h}]}(\lambda, \lambda') =  \tilde{a}^{[\mathbf{p}, \mathbf{h}]}(\lambda') \left(  K (\lambda - \lambda') + \pint_{-\infty}^{\infty} {\rm d}\alpha L^{[\mathbf{p}, \mathbf{h}]}(\lambda, \alpha)  K (\alpha, \lambda') \right). \label{Lph_int_eq}
\end{equation}
Like the resolvent dressed all the single particle eignevalues of the conserved charges, equation \eqref{eqdressing}, the generalized resolvent does the same but in the context of form factors. 

We are now finally in position to show the formula for the form factors of the density operator $\hat{\rho}(x)$ acting in position $x=0$ in the thermodynamic limit
\begin{equation} \label{FF1}
  |\langle \vartheta| \hat{\rho}(0)|\vartheta, \mathbf{h} \to \mathbf{p} \rangle | = \mathcal{A}(\vartheta, \phset) \mathcal{D}(\vartheta, \phset) \exp\left(\mathcal{B}(\vartheta, \phset)\right) ,
\end{equation}
where
\begin{align}
 \mathcal{A}(\vartheta, \phset) =& \prod_{k=1}^m \left[\frac{ 2\pi F(h_k)}{\left(k'(p_k)k'(h_k) \right)^{1/2}} \frac{\pi \tilde{F}(p_k)}{\sin \pi\tilde{F}(p_k)} \frac{\sin \pi\tilde{F}(h_k)}{\pi \tilde{F}(h_k)} \right] \nn\\
 &\times  \prod_{i,j=1}^m \left[\frac{(p_i - h_j + ic)^2}{(h_{i,j} + ic)(p_{i,j}+ic)} \right]^{1/2}\frac{\prod_{i<j=1}^m h_{i,j} p_{i,j}}{\prod_{i,j=1}^m (p_i-h_j)},\label{FF_A} \\
 \mathcal{B}(\vartheta, \phset) =&  -\frac{1}{4}\int_{-\infty}^{+\infty} {\rm d}\lambda {\rm d}\lambda' \left(\frac{\tilde{F}(\lambda) - \tilde{F}(\lambda')}{\lambda -\lambda'} \right)^2 - \frac{1}{2} \int_{-\infty}^{+\infty} {\rm d}\lambda {\rm d}\lambda' \left(\frac{\tilde{F}(\lambda)\tilde{F}(\lambda')}{(\lambda-\lambda'+ic)^2} \right)\nn\\
 &+ \sum_{k=1}^m \pint_{-\infty}^{+\infty} {\rm d}\lambda \frac{\tilde{F}(\lambda)(h_k-p_k)}{(\lambda-h_k)(\lambda-p_k)} + \int_{-\infty}^{+\infty} {\rm d}\lambda \frac{\tilde{F}(\lambda)(p_k-h_k)}{(\lambda-h_k+ic)(\lambda-p_k+ic)} \nn\\
 &+ \frac{1}{2} \delta S[\vartheta;\phset] + \frac{1}{2}\int {\rm d}\lambda\, \vartheta(\lambda)  {F}'(\lambda) \pi   {F}(\lambda) \cot(\pi  {F}(\lambda)),\label{FF_B} \\
 \mathcal{D}(\vartheta, \phset) =& \,\frac{ {\rm det}_{i,j=1}^m\left( A_{ij} + B_{ij}\right)  }{{\rm Det}(\boldsymbol{1} + L^{[\mathbf{p}, \mathbf{h}]})  {\rm Det}\Big(\boldsymbol{1}- \hat{K}_{\vartheta}\Big)}\label{FF_D}.
\end{align}
with the dressed back-flow $\tilde{F}(\lambda) = \vartheta(\lambda) F(\lambda)$. The matrix elements $A_{ij}$, $B_{ij}$ can be written in terms of ``generalized particle-hole thermodynamic functions''
\begin{align}
  A_{ij} &=  \delta_{ij}  - \frac{\resatilde^{[\mathbf{p}, \mathbf{h}]}}{\vartheta(h_i)F(h_i)}\left[ \lim_{\lambda \to h_j} \frac{L^{[\mathbf{p}, \mathbf{h}]}(h_i,\lambda)}{\tilde{a}^{[\mathbf{p}, \mathbf{h}]}(\lambda)} \right], \quad \quad \det A_{ij} = 0\\
  B_{ij} &= \frac{\resatilde^{[\mathbf{p}, \mathbf{h}]}}{\vartheta(h_i) F(h_i)} k^{\prime,[\mathbf{p}, \mathbf{h}]}(h_i) k^{\prime,[\mathbf{p}, \mathbf{h}]}(h_j).
\end{align}
with 
\begin{equation}
\vec{k}^{\prime,[\mathbf{p}, \mathbf{h}]} =  \left(\boldsymbol{1}+L^{[\mathbf{p}, \mathbf{h}]}\right) \vec{1}
\end{equation}
A more extensive explanation of expression \eqref{FF1} is provided in \cite{thermo_us}. We here focus now on the single particle-hole form factor, which constitute the leading contribution to the DSF in the small momentum limit 
\begin{align} \label{singlep_h_FF}
  |\langle \vartheta| \hat{\rho}(0) &|\vartheta, h \to p \rangle | =       \frac{  {k^{\prime,[p,h]}(h)  k^{\prime,[p,h]}(h)}{}}{\sqrt{k'(p)k'(h)   }       }   
   \left(  \frac{\pi \tilde{F}(p)}{\sin \pi\tilde{F}(p)} \frac{\sin \pi\tilde{F}(h)}{\pi \tilde{F}(h)}    \right) \frac{\sin[ \pi \tilde{F}(h)]}{\vartheta(h) \sin [ \pi F(h)]}   \nn \\& \times
e^{ - \frac{c}{2} \pint_{-\infty}^{\infty} {\rm d}
    \lambda'  \frac{\vartheta(\lambda')F(\lambda') K(\lambda' - h)}{\lambda' - h} }\exp\left(\mathcal{B}(\vartheta, [p,h])\right)  \frac{{\rm Det}\Big(\boldsymbol{1} - K^{[p,h]}\Big)  }{ {\rm Det}\Big(\boldsymbol{1}- \hat{K}_{\vartheta} \Big)} ,
\end{align} 
and its limit $p \to h$ is remarkably simple 
\begin{equation}\label{singlephlimitk0}
  |\langle \vartheta| \hat{\rho} (0)|\vartheta, h \to p \rangle |  = k^{\prime}(h)e^{\pint_{-\infty}^{+\infty} {\rm d}\lambda \frac{\tilde{F}(\lambda)(h-p)}{(\lambda-h)(\lambda-p)}} + \mathcal{O}(p-h) .
\end{equation}
Notice that if the distribution $\vartheta(\lambda)$ is smooth then the contribution given by the exponential is also of order $p-h$ and therefore it can be neglected in the small momentum limit.
\begin{equation}\label{singlephlimitk0}
  |\langle \vartheta| \hat{\rho} (0)|\vartheta, h \to p \rangle |  = k^{\prime}(h) + \mathcal{O}(p-h) .
\end{equation}
On the other hand if the distribution has a discontinuity at $\lambda = \lambda_0$, the form factor has a pole or a zero whenever $p$ or $h$ are chosen to be close to $\lambda_0$. In this case then we can write the small momentum limit by isolating the divergent part and including the regular part in the higher order corrections 
\begin{equation}\label{singlephlimitk02}
  |\langle \vartheta| \hat{\rho} (0)|\vartheta, h \to p \rangle |  = k^{\prime}(h) \left|\frac{\lambda_0 - h}{\lambda_0 - p}\right|^{\Delta\vartheta F(\lambda_0)} + \mathcal{O}(p-h) ,
\end{equation}
where $\Delta\vartheta = \lim_{\epsilon\rightarrow 0} \vartheta(\lambda_0 +\epsilon) - \vartheta(\lambda_0 - \epsilon)$ is the height of the discontinuity.

\subsection{Large coupling expansion}

At large values of $c$ the expression for the form factors drastically simplifies. We obtain 
\begin{equation}\label{eq:largec}
 |\langle \vartheta| \hat{\rho} (0)|\vartheta, h \to p \rangle |  = (1 + 2n/c)   e^{\pint_{-\infty}^{+\infty} {\rm d}\lambda \frac{\tilde{F}(\lambda)(h-p)}{(\lambda-h)(\lambda-p)}} + O(1/c^2)
\end{equation}
for the single particle-hole form factor and its leading part is of order $1$.  The two particle hole form factors are proportional to $1/c$, and therefore their contribution to the DSF scales as $1/c^2$. Notice that the expression \eqref{eq:largec} reproduces the result of \cite{PhysRevA.73.023612} for the thermal case. Fig.~\ref{fig:DSF} shows the DSF computed with form factors~\eqref{eq:largec} in the NESS state and at finite temperature.


\begin{figure}[h]
 \centering
 \includegraphics[width=0.5\hsize]{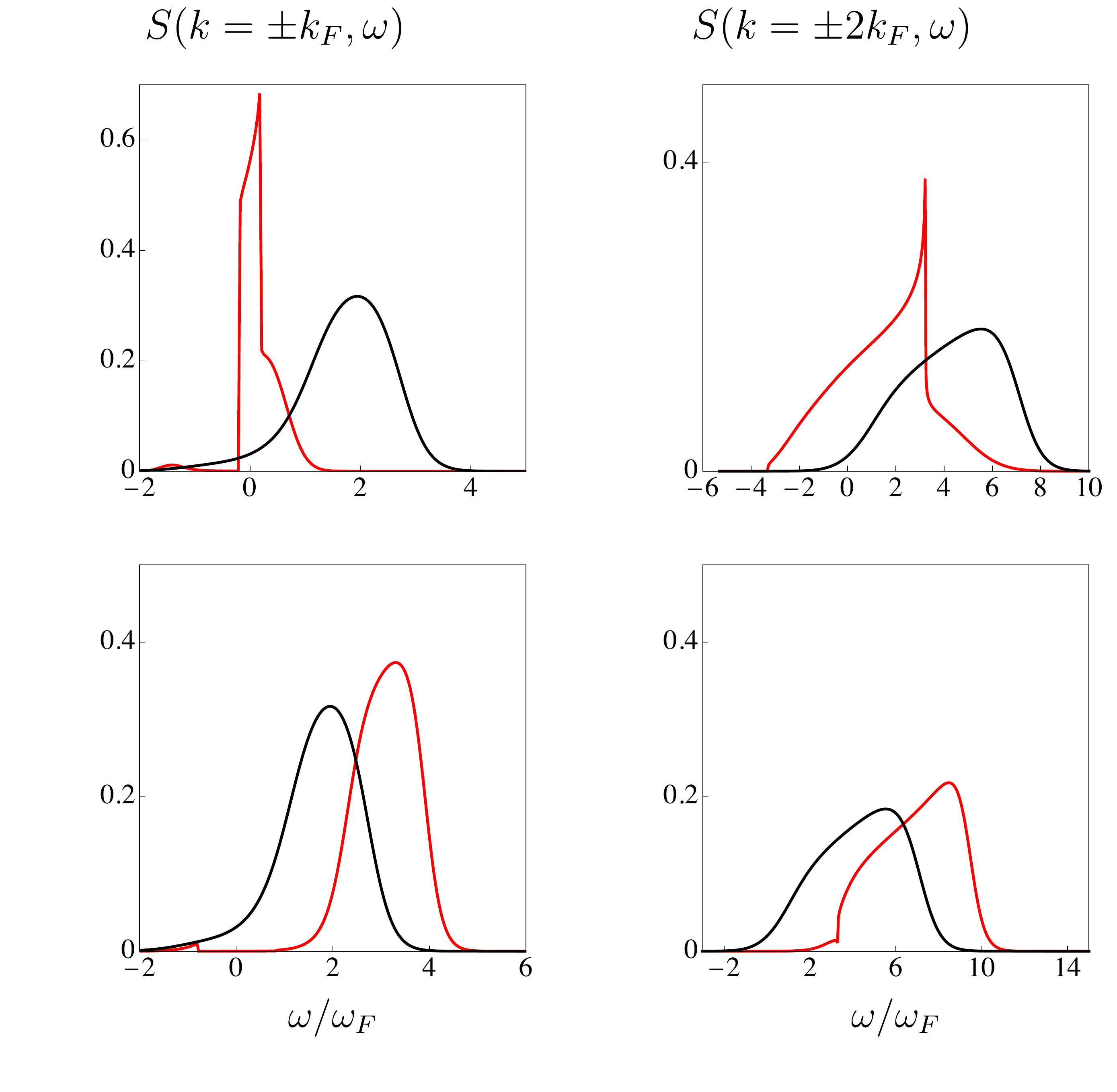} 
\caption{\textsc{dsf} of the \textsc{ness} (red line) and on a thermal state with $T=1$, $n=n_{\text{\textsc{ness}}}$, and $c=10$. Plots on top display data with $k=k_F$ and the ones on the bottom with $k=-k_F$ ($k_F = \pi n $). The \textsc{ness} is obtained by joining a left gas with $T_L=1,n_L=1$ and a right gas with $T_R=1,n_R= 0.1$. The coupling strength $c=10$ of both gases is the same  (density $n_{\text{\textsc{ness}}} = 0.54$).  As the coupling $c$ is relatevely large, the \textsc{dsf} is well approximated by the single particle-hole contribution. } \label{fig:DSF}
 \end{figure}

\end{document}